\documentclass[prc,amsmath,amssymb,twocolumn,preprintnumbers,longbibliography]{revtex4-2}
\usepackage{amsmath, amsthm, amssymb, braket}

\usepackage{amsfonts}

\usepackage{hyperref}
\usepackage[dvipdfmx]{graphicx}
\usepackage{color}
\usepackage{dcolumn}
\usepackage{bm}

\newcommand{\be}{\begin{equation}}
\newcommand{\ee}{\end{equation}}
\newcommand{\dd}{\mathrm{d}}
\newcommand{\br}{\boldsymbol{r}}

\begin{document}
\allowdisplaybreaks[1]

\title{Pairing and nonaxial-shape correlations in $N=150$ isotones 
}

\author{Kenichi Yoshida}
\email[E-mail: ]{kyoshida@ruby.scphys.kyoto-u.ac.jp}
\affiliation{Department of Physics, Kyoto University, Kyoto, 606-8502, Japan}

\preprint{KUNS-2874}
\date{\today}

\begin{abstract}
\begin{description}
\item[Background] 
Rotational bands have been measured around $^{250}$Fm associated with strong deformed-shell closures. 
The $K^\pi=2^-$ excited band emerges systematically in $N=150$ isotones raging from plutonium to nobelium with even-$Z$ numbers, 
and a sharp drop in energies was observed in californium.
\item[Purpose] I attempt to uncover the microscopic mechanism for the appearance of such a low-energy $2^-$ state in $^{248}$Cf. 
Furthermore, I investigate the possible occurrence of the low-energy $K^\pi=2^+$ state, the $\gamma$ vibration, to elucidate the mechanism 
that prefers the simultaneous breaking of the reflection and axial symmetry to the breaking of the axial symmetry alone in this mass region. 
\item[Method] I employ a nuclear energy-density functional (EDF) method: the Skyrme--Kohn--Sham--Bogoliubov and the quasiparticle 
random-phase approximation are used to describe the ground state and the transition to excited states. 
\item[Results] The Skyrme-type SkM* and SLy4 functionals reproduce the fall in energy, but not the absolute value, of the $K^\pi=2^-$ state at $Z=98$, 
where the proton two-quasiparticle (2qp) excitation $[633]7/2 \otimes [521]3/2$ plays a decisive role for the peculiar isotonic dependence. 
I find interweaving roles by the pairing correlation of protons and the deformed shell closure at $Z=98$. 
The SkM* model predicts the $K^\pi=2^-$ state appears lower in energy in $^{246}$Cf than in $^{248}$Cf as the Fermi level of neutrons is located in between the  
$[622]5/2$ and $[734]9/2$ orbitals.
Except for $^{250}$Fm in the SkM* calculation, 
the $K^\pi=2^+$ state is predicted to appear higher in energy than the $K^\pi=2^-$ state 
because the quasi-proton $[521]1/2$ orbital is located above the $[633]7/2$ orbital.
\item[Conclusions]
A systematic study of low-lying collective states in heavy actinide nuclei 
provides a rigorous testing ground for microscopic nuclear models.
The present study shows a need for improvements in the EDFs to describe pairing correlations and shell structures in heavy nuclei, 
that are indispensable in predicting the heaviest nuclei. 
\end{description}
\end{abstract}

\maketitle

\section{Introduction}\label{intro}

Recent developments in experimental technology and theoretical modeling have significantly advanced the study of exotic nuclei. 
The existence of superheavy nuclei in the limit of a great number of protons has long been a subject of interest in nuclear physics~\cite{sob07,oga15,ada21}. 
The shell effect is a key to define their stability.
The pairing correlations further play an influential role in determining the stability and shape of the ground state, 
and accordingly, the height of the fission barrier~\cite{kar10}.
However, it is not simple to investigate the shell structure and the pairing in such nuclei in the extreme regime. 
Strongly deformed actinide nuclei have thus been investigated 
as they can reveal the nuclear-structure information in the island of enhanced stability of spherical superheavy nuclei~\cite{ack17}.

Details of the spectroscopic information in the actinide nuclei 
have provided an excellent testing ground for the reliability and predicting power of modern nuclear-structure theories~\cite{her08,the15}. 
Several deformed shell closures are expected to emerge at $Z=98$ and 100 and at $N=150$ and 152 in the actinide nuclei~\cite{ben03b}. 
Correspondingly, beautiful rotational bands have been measured~\cite{her08,the15}, that are indicative of an axially-deformed rotor. 
The rotational properties and the high-$K$ isomers have been studied in detail in terms of the pairing~\cite{dug01,afn03,tan06,gre12,afn13}. 

In the course of the studies, anomalies have been revealed that are caused by residual interactions beyond the mean-field picture. 
A collective phenomenon shows up as a low-lying $K^\pi=2^-$ state in the $N=150$ isotones, 
and the energy falls peculiarly in $^{248}$Cf~\cite{yat75,yat75b}.
The nonaxial reflection-asymmetric $\beta_{32}$ (dynamic) fluctuation and (static) deformation 
have been suggested by the projected shell-model calculation~\cite{che08} 
and the relativistic energy-density-functional (EDF) calculation~\cite{zha12}. 
The quasiparticle-random-phase approximation (QRPA) calculations 
have also been performed to investigate the vibrational character~\cite{rob08,rez18}. 
While the calculation using the Nilsson potential describes the anomalous behavior in $^{248}$Cf, 
the selfconsistent QRPA calculation employing the Gogny-D1M EDF shows a smooth isotonic dependence. 

The observation of a strong $\beta_{32}$ correlation that breaks the axial symmetry 
suggests that the non-axial quadrupole deformation, $\gamma$ deformation, may also occur.
Nuclear triaxiality brings about exotic collective modes: 
the appearance of the low-energy $K^\pi = 2^+$ band, the $\gamma$ band, is a good indicator for the $\gamma$ deformation~\cite{BM2,rin80}. 
The $\gamma$ vibrational mode of excitation is regarded as a precursory soft mode of the permanent $\gamma$ deformation. 
Experimentally, the $K^\pi=2^+$ state has not been observed to be as low in energy as the $K^\pi=2^-$ state in $^{244}$Pu and $^{246}$Cm~\cite{mul71,tho75}. 
Therefore, it is interesting to investigate a microscopic mechanism that prefers the simultaneous breaking of the reflection and axial symmetry to the breaking of the axial symmetry alone. 

Pairing correlations are essential for describing low-energy vibrational modes of excitation~\cite{mat13} 
and are a key to understanding the collective nature. 
Therefore, in this article, I am going to investigate the role of the pairing correlations in both the static and dynamic aspects, 
as done in the studies in exotic nuclei in the neutron-rich region~\cite{yos06b,yos06,yos08b}. 
To this end, I employ a nuclear EDF method in which the pairing and deformation are simultaneously considered. 
Since the nuclear EDF method is a theoretical model being capable of handling nuclides with arbitrary mass numbers in a single framework~\cite{ben03,nak16}, 
the present investigation in the actinides can give an insight into the veiled nuclear structure of superheavy nuclei.

This paper is organized in the following way: 
the theoretical framework for describing the low-lying vibrations is given in Sec.~\ref{model} and 
details of the numerical calculation are also given; 
Sec.~\ref{result} is devoted to the numerical results and discussion based on the model calculation; 
after the discussion on the ground-state properties: the static aspects of pairing and deformation in Sec.~\ref{GS}, 
the discussion on the $K^\pi=2^-$  and $K^\pi=2^+$ states is given in Sec.~\ref{oct2_mode} and Sec.~\ref{gam_mode}, respectively; 
then, a summary is given in Sec.~\ref{summary}.

\section{Theoretical model}\label{model}

\subsection{KSB and QRPA calculations}

Since the details of the framework can be found in Refs.~\cite{yos08,yos13b}, 
here I briefly recapitulate the basic equations relevant to the present study. 
In the framework of the nuclear EDF method I employ, 
the ground state is described by solving the 
Kohn--Sham--Bogoliubov (KSB) equation~\cite{dob84}:
\begin{align}
\sum_{\sigma^\prime}
\begin{bmatrix}
h^q_{\sigma \sigma^\prime}(\br)-\lambda^{q}\delta_{\sigma \sigma^\prime} & \tilde{h}^q_{\sigma \sigma^\prime}(\br) \\
\tilde{h}^q_{\sigma \sigma^\prime}(\br) & -h^q_{\sigma \sigma^\prime}(\br)+\lambda^q\delta_{\sigma \sigma^\prime}
\end{bmatrix}
\begin{bmatrix}
\varphi^{q}_{1,\alpha}(\br \sigma^\prime) \\
\varphi^{q}_{2,\alpha}(\br \sigma^\prime)
\end{bmatrix} \notag \\
= E_{\alpha}
\begin{bmatrix}
\varphi^{q}_{1,\alpha}(\br \sigma) \\
\varphi^{q}_{2,\alpha}(\br \sigma)
\end{bmatrix}, \label{HFB_eq}
\end{align}
where 
the single-particle and pair Hamiltonians $h^q_{\sigma \sigma^\prime}(\br)$ and $\tilde{h}^q_{\sigma \sigma^\prime}(\br)$ are given by the functional derivative of the EDF 
with respect to the particle density and the pair density, respectively. 
An explicit expression of the Hamiltonians is found in the Appendix of Ref.~\cite{kas21}. 
The superscript $q$ denotes 
$\nu$ (neutron, $ t_z= 1/2$) or $\pi$ (proton, $t_z =-1/2$). 
The average particle number is fixed at the desired value by adjusting the chemical potential $\lambda^q$. 
Assuming the system is axially symmetric, 
the KSB equation (\ref{HFB_eq}) is block diagonalized 
according to the quantum number $\Omega$, the $z$-component of the angular momentum. 

The excited states $| i \rangle$ are described as 
one-phonon excitations built on the ground state $|0\rangle$ as 
\begin{align}
| i \rangle &= \hat{\Gamma}^\dagger_i |0 \rangle, \\
\hat{\Gamma}^\dagger_i &= \sum_{\alpha \beta}\left\{
f_{\alpha \beta}^i \hat{a}^\dagger_{\alpha}\hat{a}^\dagger_{\beta}
-g_{\alpha \beta}^i \hat{a}_{\bar{\beta}}\hat{a}_{\bar{\alpha}}\right\},
\end{align}
where $\hat{a}^\dagger$ and $\hat{a}$ are 
the quasiparticle (qp) creation and annihilation operators that 
are defined in terms of the solutions of the KSB equation (\ref{HFB_eq}) with the Bogoliubov transformation.
The phonon states, the amplitudes $f^i, g^i$ and the vibrational frequency $\omega_i$, 
are obtained in the quasiparticle-random-phase approximation (QRPA): the linearized time-dependent density-functional theory for superfluid systems~\cite{nak16}.  
The EDF gives the residual interactions entering into the QRPA equation. 
In the present calculation scheme, the QRPA equation 
is block diagonalized according to the quantum number $K=\Omega_\alpha + \Omega_\beta$. 

\subsection{Numerical procedures}

I solve the KSB equation in the coordinate space using cylindrical coordinates
$\boldsymbol{r}=(\varrho,z,\phi)$.
Since I assume further the reflection symmetry, only the region of $z\geq 0$ is considered. 
I use a two-dimensional lattice mesh with 
$\varrho_i=(i-1/2)h$, $z_j=(j-1)h$ ($i,j=1,2,\dots$) 
with a mesh size of
$h=0.6$ fm and 25 points for each direction. 
The qp states are truncated according to the qp 
energy cutoff at 60 MeV, and 
the qp states up to the magnetic quantum number $\Omega=23/2$
with positive and negative parities are included. 
I introduce the truncation for the two-quasiparticle (2qp) configurations in the QRPA calculations,
in terms of the 2qp-energy as 60 MeV. 

For the normal (particle--hole) part of the EDF,
I employ mainly the SkM* functional~\cite{bar82}, and use the SLy4 functional~\cite{cha98} to complement the discussion. 
For the pairing energy, I adopt the so-called mixed-type interaction:
\be
V_{\rm{pair}}^{q}(\br,\br^\prime)=V_0^{q}
\left[ 1-\frac{\rho(\br)}{2\rho_0} \right]
\delta(\br-\br^\prime)
\ee
with $\rho_0=0.16$ fm$^{-3}$, and $\rho(\br)$ being the isoscalar (matter) particle density. 
The same pair interaction is employed for the dynamical pairing 
in the QRPA calculation. 
The parameter $V_0^q$ is fitted to the three-point formula for the odd-even staggering
centered at the odd-mass system and averaged over the two neighboring nuclei~\cite{sat98,dob02}. 
In the present work, I fix the pairing parameters to reproduce reasonably the data of $^{244}$Cm 
which are 0.63 MeV and 0.57 MeV for neutrons and protons, respectively.
The resultant pairing gaps are 0.65 MeV and 0.59 MeV with $V_0^\nu=-270$ MeV fm$^3$ and $V_0^\pi=-310$ MeV fm$^3$.

\section{results and discussion}\label{result}
\subsection{Ground-state properties}\label{GS}

\begin{table}[t]
\caption{\label{tab:def} 
Calculated deformation parameters $\beta_2$. 
The evaluated data denoted as exp. are taken from Ref.~\cite{pri16}. 
Listed in parentheses are the intrinsic quadrupole moments in the unit of $e$b and b for protons and neutrons, respectively.}
\begin{ruledtabular}
\begin{tabular}{lcccccc}
& \multicolumn{3}{c}{protons} & & \multicolumn{2}{c}{neutrons}  \\ 
\cline{2-4} \cline{6-7}
& SkM* & SLy4 & exp. & & SkM* & SLy4 \\
\hline
$^{244}$Pu & 0.29 (11.7) & 0.28 (11.5)  & 0.29 && 0.27 (18.3) & 0.27 (18.2)  \\
$^{246}$Cm & 0.29 (12.3) & 0.28 (12.2) & 0.30 && 0.27 (18.7) & 0.27 (18.6) \\
$^{248}$Cf & 0.30 (12.9) & 0.30 (12.9) & && 0.27 (19.0) & 0.27 (19.0)  \\
$^{250}$Fm & 0.30 (13.5) & 0.30 (13.4) & && 0.28 (19.3) & 0.28 (19.2) \\
$^{252}$No & 0.30 (13.9) & 0.30 (13.7) & && 0.28 (19.4) & 0.28 (19.2)
\end{tabular}
\end{ruledtabular}
\end{table}

Table~\ref{tab:def} summarizes the calculated deformation parameters:
\be
\beta_2^q=\dfrac{4\pi}{5N_q \langle r^2\rangle_q}\int \dd \br Y_{20}(\hat{r})\rho_q(\br),
\ee
where the rms radius is given as $\sqrt{\langle r^2\rangle_q}=\sqrt{\int r^2\rho_q(\br) \dd \br/N_q}$, 
with $N_q$ being either the neutron number or proton number.
They are compared with the experimental data~\cite{pri16}, which is evaluated from the $B({\rm E2})$ value 
based on the leading-order intensity relations of E2 matrix elements for an axially symmetric rotor~\cite{BM2}. 
The calculated results obtained by employing the SLy4 functional are also included. 
The strengths of the pair interaction were fitted similarly for SkM*; 
the strength $V_0^\nu=-310$ MeV fm$^3$ and $V_0^\pi=-320$ MeV fm$^3$ produces the 
pair gap as 0.67 MeV and 0.53 MeV for neutrons and protons, respectively. 
Both the SkM* and SLy4 functionals reproduce well the strong deformation of $^{244}$Pu and $^{246}$Cm. 
The calculations predict a stronger deformation for higher $Z$ isotones: $^{250}$Fm and $^{252}$No. 

\begin{figure}[t]
\includegraphics[scale=0.4]{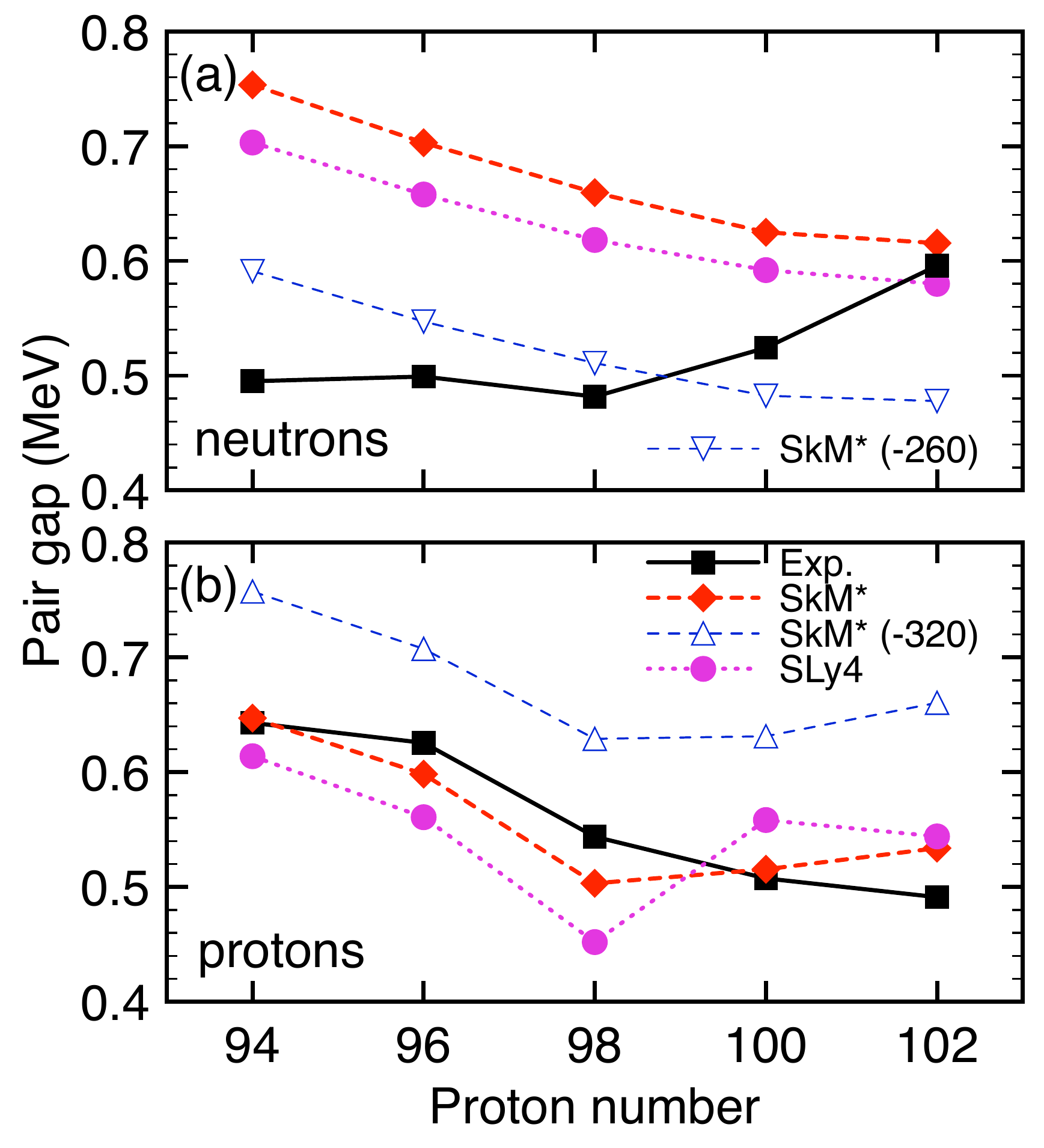}
\caption{\label{fig:gap}
Pair gaps of neutrons and protons in $^{244}$Pu, $^{246}$Cm, $^{248}$Cf, $^{250}$Fm, and $^{252}$No. 
The calculated gaps (filled diamonds) are compared with the experimental data (filled squares). 
The results obtained by reducing the pairing strength $V_0^\nu=-260$ MeV fm$^3$ while keeping the strength for protons,  
and those obtained by increasing the pairing strength $V_0^\pi=-320$ MeV fm$^3$ while keeping the strength for neutrons 
are also shown by open triangles for neutrons and protons, respectively. 
Depicted also are the results obtained by employing the SLy4 functional (filled circles). 
 }
\end{figure}

I show in Fig.~\ref{fig:gap} the pair gaps of protons and neutrons:
\be
\Delta^q = \left| \dfrac{\int \dd \br \tilde{\rho}_q(\br) \tilde{U}^q(\br)}{\int \dd \br \tilde{\rho}_q(\br)} \right|,
\ee
where $\tilde{U}^q(\br)$ is the pair potential given by 
$\tilde{h}^q_{\sigma \sigma^\prime}(\br)=\delta_{\sigma \sigma^\prime}\tilde{U}^q(\br)$
and $\tilde{\rho}_q(\br)$ is the anomalous (pair) density. 
The calculations overestimate the measurements for neutrons. 
When I slightly decrease the pair strength for neutrons as $V_0^\nu=-260$ MeV fm$^3$, 
the calculation shows a reasonable agreement. 
However, a proton-number dependence is opposite to what is shown in the measurements. 
This indicates that 
the neutrons deformed gap of 150 must be more significant than that of 148 in lower $Z$ and that 
the level density around $N=150$ is high in higher $Z$ nuclei. 
Since the deformed shell gap at 150 is formed by the up-sloping orbitals stemming from the $2g_{9/2}$ shell 
and the down-sloping orbitals emanating from the $2g_{7/2}$ shell, 
a slight modification of the spin--orbit interaction would impact the shell structure in this region, 
and further, the prediction of the magic numbers in the superheavy region, as anticipated in Ref.~\cite{ben03b}. 

\begin{figure}[t]
\begin{center}
\includegraphics[scale=0.76]{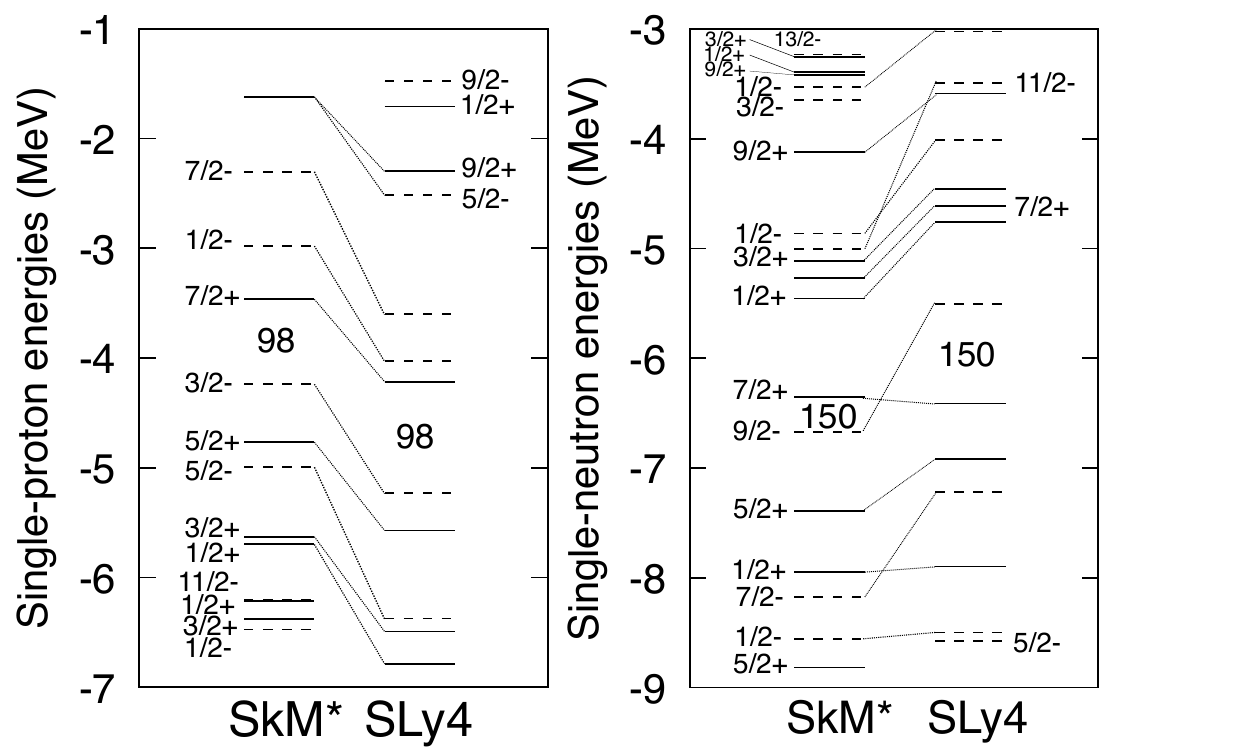}
\caption{\label{fig:shell} 
Single-particle energies of protons and neutrons near the Fermi levels in $^{248}$Cf. 
The results obtained by using the SkM* and SLy4 functionals are compared. 
The solid and dashed lines indicate the positive- and negative-parity states, respectively. 
Each orbital is labeled by $\Omega^\pi$. 
}
\end{center}
\end{figure}

The resultant pair gaps of protons are compatible with the experimental ones. 
The SkM* functional with the mixed-type pairing well reproduces the overall isotonic dependence, 
while the SLy4 shows a drop at $Z=98$, which the measurements do not reveal. 
One may conjecture that this is because the deformed shell gap is overdeveloped at $Z=98$ in the SLy4 model. 
I then show in Fig.~\ref{fig:shell} the single-particle levels in $^{248}$Cf to visualize the shell structure. 
The single-particle orbitals were obtained by rediagonalizing the single-particle Hamiltonian:
\be
\sum_{\sigma^\prime} h^q_{\sigma \sigma^\prime}[\rho_\ast,\tilde{\rho}_\ast]
\varphi_i^q(\br \sigma^\prime)=\varepsilon_i^q \varphi_i^q(\br \sigma)
\ee
with $\rho_\ast$ and $\tilde{\rho}_\ast$ obtained as the solution of the KSB equation (\ref{HFB_eq}). 
The energy gap at $Z=98$ is indeed larger than that obtained by using the SkM* functional, 
as discussed in Ref.~\cite{cha06}. 
Since the pair gap is about 0.5 MeV, the orbitals with $\Omega^\pi=7/2^+$ and $3/2^-$ are primarily active 
for the pairing correlation. The occupation probability of the orbital with $\Omega^\pi=7/2^+$ is indeed only 0.14, 
and that of the orbital with $\Omega^\pi=3/2^-$ is as large as 0.85. 
For reference, they are 0.23 and 0.80 in the SkM* model. 
A strong deformed-shell closure at $Z=98$ in the SLy4 model 
is unfavorable in generating the low-energy $K^\pi=2^-$ state, as questioned in Ref.~\cite{rob08}. 

\subsection{Vibrational states}\label{vib}

I consider the response to 
the quadrupole ($\lambda=2$) and octupole ($\lambda=3$) operators defined by
\begin{align}
	\hat{F}^q_{\lambda K}
	=& \sum_{\sigma}\int \dd \br r^\lambda Y_{\lambda K}
	\hat{\psi}^\dagger_q(\br \sigma)\hat{\psi}_q(\br \sigma), \label{oct_op}
\end{align}
where $\hat{\psi}^\dagger_q(\br \sigma), \hat{\psi}_q(\br \sigma)$ represent the nucleon field operators. 
The reduced transition probabilities can be evaluated with the intrinsic transition strengths as 
\begin{align}
	B({\rm E\lambda};I^\pi_K\to 0^+_{\rm gs})&=\frac{2-\delta_{K,0}}{2I_K +1} e^2| \langle i |\hat{F}^\pi_{\lambda K} |0 \rangle |^2, \\
	B({\rm IS\lambda};I^\pi_K\to 0^+_{\rm gs})&=\frac{2-\delta_{K,0}}{2I_K +1} | \langle i |\hat{F}^\pi_{\lambda K}+\hat{F}^\nu_{\lambda K} |0 \rangle |^2
\end{align}
in the rotational coupling scheme~\cite{BM2}, where $I_K$ represents the nuclear spin of the $K^\pi$-band. 
The intrinsic transition matrix element is given as
\begin{align}
\langle i | \hat{F}^q_{\lambda K}|0 \rangle &= \sum_{\alpha \beta} (f^i_{\alpha \beta}+g^i_{\alpha \beta}) \langle \alpha \beta |\hat{F}^q_{\lambda K}|0\rangle \\
& \equiv \sum_{\alpha \beta} M^{q,i}_{\alpha \beta} \label{eq:ME} 
\end{align}
with the QRPA amplitudes and the 2qp matrix elements. 
The QRPA amplitudes are normalized as
\begin{equation}
\sum_{\alpha \beta} \left[ |f^i_{\alpha \beta}|^2 - |g^i_{\alpha \beta}|^2 \right]=1.
\end{equation}

\subsubsection{$K^\pi=2^-$ state}\label{oct2_mode}

\begin{figure}[t]
\begin{center}
\includegraphics[scale=0.4]{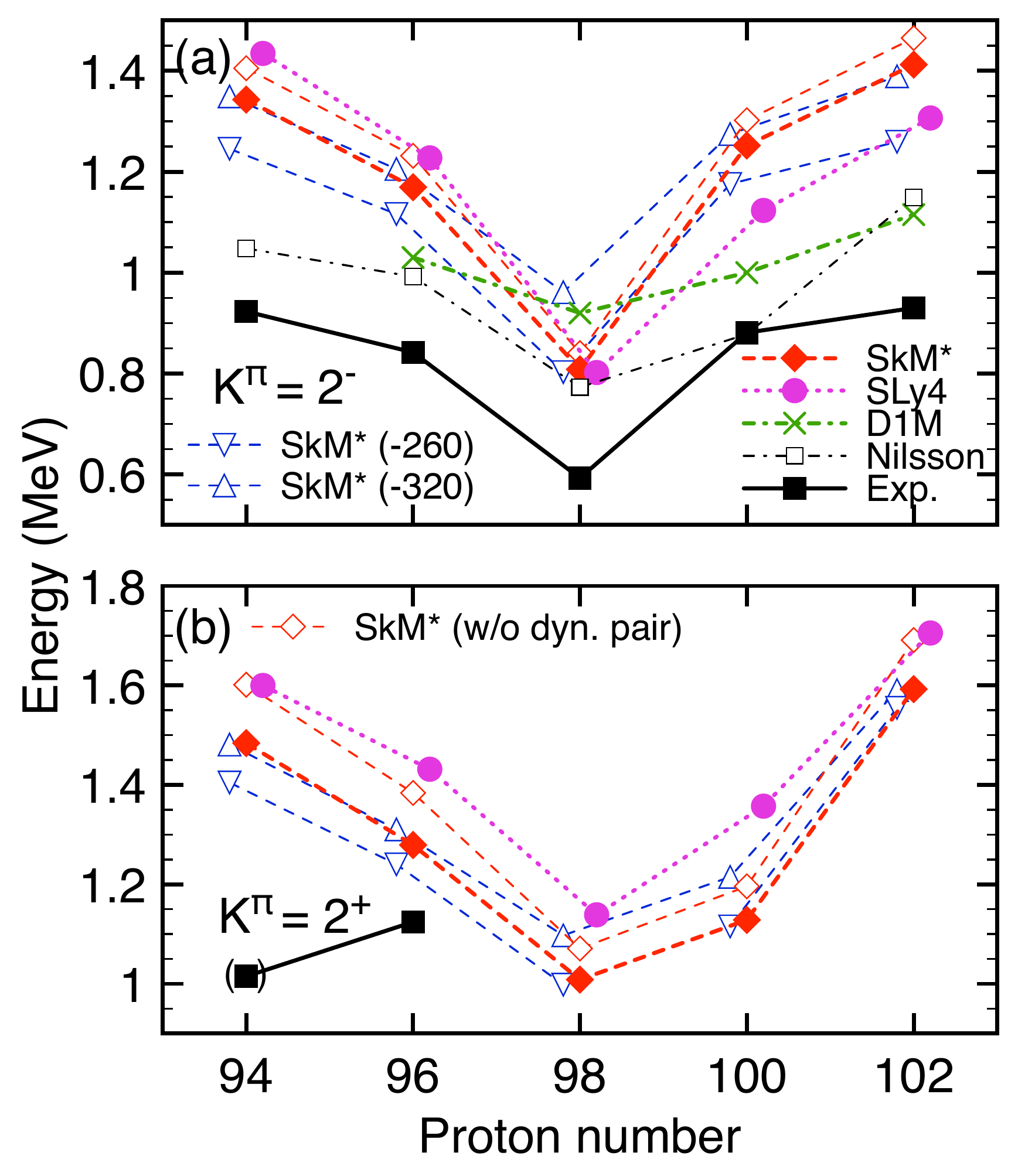}
\caption{\label{fig:QRPA2} 
Excitation energies of the (a) $K^\pi=2^-$ and (b) $K^\pi=2^+$ states in the $N=150$ isotones. 
The calculations employing the SkM* and SLy4 functionals are compared with the measurements~\cite{rob08,mul71,tho75}. 
The results obtained by increasing the pairing for protons ($V_0^\pi=-320$ MeV fm$^3$) 
and decreasing the pairing for neutrons ($V_0^\nu=-260$ MeV fm$^3$) are depicted by open triangles. 
The SkM* results without the dynamical pairing are given by open diamonds. 
Shown is also the calculations using the Gogny D1M functional~\cite{rez18} and the Nilsson potential~\cite{rob08}.
}
\end{center}
\end{figure}

I show in Fig.~\ref{fig:QRPA2}(a) the intrinsic excitation energies of the $K^\pi=2^-$ state in the $N=150$ isotones. 
Both the SkM* and SLy4 functionals well reproduce the isotonic trend of the energy, particularly a drop at $Z=98$, 
but overestimate the measurements as a whole. 
The QRPA calculation in Ref.~\cite{rob08} based on the Nilsson potential~\cite{nak96} also well describes a sharp fall 
in energies at $Z=98$, while the one employing the D1M functional shows only a smooth proton-number dependence. 
It is noted that in these calculations, the rotational correction and the Coriolis coupling effect~\cite{nee70b} are not included. 
One needs to consider these effects for a quantitative description, but it is beyond the scope of the present study. 
Rather, I am going to investigate a microscopic mechanism for the unique isotonic dependence of the octupole correlation. 

A lowering in energy implies that the collectivity develops from $Z=96$ to $Z=98$. 
However, the isoscalar intrinsic transition strengths to the $K^\pi=2^-$ state do not show such a trend: 
6.83, 6.18, 6.11, 4.74, and 4.45 (in $10^5$ fm$^6$) in $^{244}$Pu, $^{246}$Cm, $^{248}$Cf, $^{250}$Fm, and $^{252}$No. 
The isotonic dependence of the properties of the $K^\pi=2^-$ state is actually governed by the details around the Fermi surface: 
the interplay between the underlying shell structures and the pairing correlations. 
To see the role of the pairing, shown in Fig.~\ref{fig:QRPA2}(a) are the results obtained by varying the pair interaction. 
I reduced the pairing strength $V_0^\nu=-260$ MeV fm$^3$ while keeping the strength for protons, 
and I increased the pairing strength $V_0^\pi = -320$ MeV fm$^3$ while keeping the strength for neutrons. 
Then, I find that the $K^\pi=2^-$ state in $^{248}$Cf is sensitive to the pairing of protons. 
When the strength of the pair interaction for protons is increased, 
the excitation energy in $^{248}$Cf becomes higher: 0.84 MeV to 1.00 MeV, 
while the excitation energy in the other isotones does not change much. 
As seen in Fig.~\ref{fig:gap}(b), $2\Delta$ for protons 
rises by about 0.2 MeV in increasing the pairing strength for protons: 
the increase in the pairing gap of protons directly affects the increase in the excitation energy.  
On the other hand, the change in the pairing strength for neutrons does not alter the energy in $^{248}$Cf. 
This indicates that the $K^\pi=2^-$ state is sensitive to the excitation of protons.  
The calculation in Ref.~\cite{rob08} found a high amplitude for the proton configuration of $[633]7/2 \otimes [521]3/2$, 
and the quasiparticle phonon model predicted that the main component of the wave function is this proton configuration with a weight of 62\%~\cite{jol11}. 
Containing a large component of this proton configuration is supported by the population of the $K^\pi=2^-$ and $5^-$ states in 
the $^{249}$Bk$(\alpha, \rm{t})$$^{250}$Cf reaction~\cite{yat76}. 

\begin{figure}[t]
\begin{center}
\includegraphics[scale=0.28]{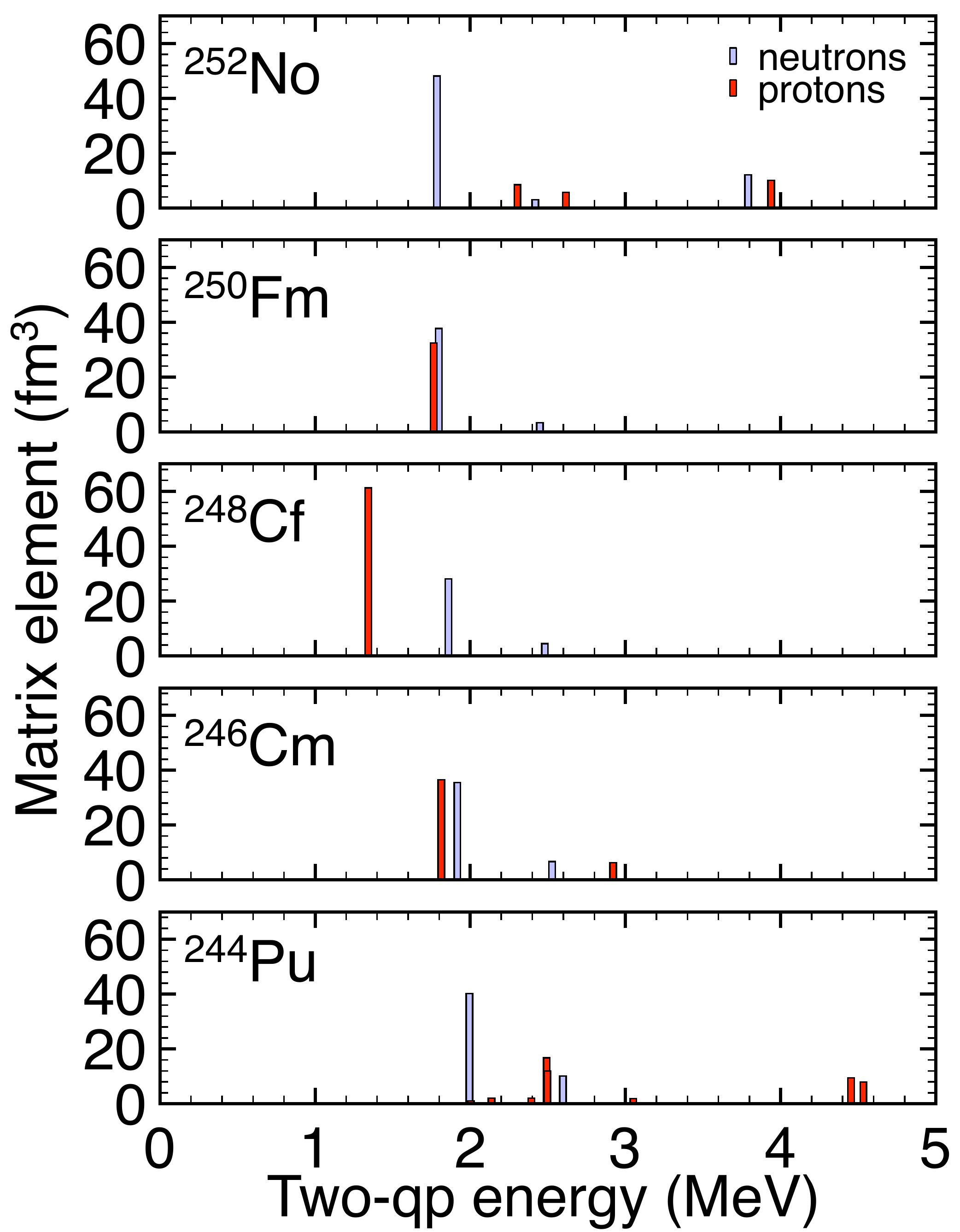}
\caption{\label{fig:ME_oct} 
Matrix elements $M_{\alpha \beta}$ (\ref{eq:ME}) of the 2qp excitations near the Fermi levels 
for the $K^\pi=2^-$ state as functions of the unperturbed 2qp energy $E_\alpha + E_\beta$.
}
\end{center}
\end{figure}

Let me investigate the isotonic evolution of the microscopic structure. 
Figure~\ref{fig:ME_oct} shows the matrix elements $M_{\alpha \beta}$ defined in Eq.~(\ref{eq:ME}) of the 2qp excitations near the Fermi levels for the $K^\pi=2^-$ state. 
One can see that 2qp excitations near the Fermi levels have a coherent contribution to generate the $K^\pi=2^-$ state with the same sign. 
It should be noted that not only the 2qp excitations near the Fermi levels but those in the giant-resonance region have 
a coherent contribution for the enhancement in the transition strength to the low-frequency mode, as discussed in Ref.~\cite{yos08b}. 
In $^{244}$Pu, the neutron 2qp excitation of $[622]5/2 \otimes [734]9/2$, whose 2qp energy is about 2 MeV, 
is a main component with an amplitude $|f|^2-|g|^2$ of 0.45. 
This is why the decrease in the pairing strength for neutrons lowers the excitation energy. 
When adding two protons to $^{244}$Pu, the Fermi level of protons becomes higher. 
The proton 2qp excitation of $[521]3/2 \otimes [633]7/2$, whose 2qp energy decreases to 1.8 MeV, 
then has an appreciable contribution with an amplitude of 0.43, 
together with the $\nu[622]5/2 \otimes \nu[734]9/2$ excitation with an amplitude of 0.31. 
Note that both the $\pi[521]3/2$ and $\pi[633]7/2$ orbitals are particle-like. 
In $^{248}$Cf, the Fermi level of protons is located just in between the $\pi[521]3/2$ and $\pi[633]7/2$ orbitals. 
Then, the 2qp excitation of $\pi[521]3/2 \otimes \pi[633]7/2$ appears low in energy at 1.3 MeV, and 
the $K^\pi=2^-$ state is predominantly generated by this configuration with an amplitude of 0.72. 
The amplitude of the $\nu[622]5/2 \otimes \nu[734]9/2$ excitation declines to 0.14. 
Adding two more protons, the 2qp excitation of $\pi[521]3/2 \otimes \pi[633]7/2$ is a hole--hole type excitation, 
and the unperturbed 2qp states are located higher in energy. 
Therefore, the matrix element and contribution of this 2qp excitation decrease. 
The $K^\pi=2^-$ state in $^{250}$Fm is mainly constructed by 
the $\pi[521]3/2 \otimes \pi[633]7/2$ and $\nu[622]5/2 \otimes \nu[734]9/2$ 
excitations with a wight of 0.42 and 0.41, respectively. 
Then, the $\nu[622]5/2 \otimes \nu[734]9/2$ excitation dominantly 
generates the $K^\pi=2^-$ state with a weight of 0.73 in $^{252}$No. 

It seems that the collectivity of the $K^\pi=2^-$ state in $^{248}$Cf is weaker than in the other isotones: 
single 2qp excitation dominates the state.  
The sum of the backward-going amplitudes $\sum|g|^2$ and the octupole polarizability $\chi$ can be measures of the correlation, 
in addition to the energy shift and the enhancement in the transition strength due to the QRPA correlations. 
The calculated $\sum|g|^2$ is 0.22, 0.20, 0.24, 0.14, and 0.12 and 
the stiffness parameter $C_{\beta_{32}}=\chi^{-1}_{\beta_{32}}$ (in MeV) is 418, 402, 306, 502, and 548 
in $^{244}$Pu, $^{246}$Cm, $^{248}$Cf, $^{250}$Fm, and $^{252}$No, respectively. 
Based on these measures, the octupole correlation is strong in $^{248}$Cf as well as in $^{244}$Pu and $^{246}$Cm.

In Fig.~\ref{fig:QRPA2}(a), I plot the calculated results obtained by neglecting the dynamical pairing, as depicted by open diamonds. 
The dynamical pairing lowers the $K^\pi=2^-$ state by 0.04--0.06 MeV. 
One sees a larger effect in $^{246}$Cm. 
The energy shift due to the dynamical pairing is 0.06 MeV out of 0.65 MeV due to the QRPA correlations in total. 
It is noted that the pairing correlation energy in the total binding energy is about 0.7\%. 
As discussed above, the collective state in $^{246}$Cm is constructed mainly by 
the $\nu[622]5/2 \otimes \nu[734]9/2$ and $\pi[521]3/2 \otimes \pi[633]7/2$ excitations. 
Since these are a neutron hole--hole and a proton particle--particle excitations, 
the residual pair interaction for both neutrons and protons is active and instrumental in forming the collective state. 

Surprisingly, the SLy4 model reproduces the drop in energies at $Z=98$. 
As mentioned above, and as questioned in Ref.~\cite{rob08}, Fig.~\ref{fig:shell} shows that the SLy4 model overestimates the deformed gap at $Z=98$, 
requiring a high energy particle--hole excitation from $\pi[521]3/2$ to $\pi[633]7/2$. 
Here, the pairing correlations are key to understanding the characteristic isotonic dependence. 
The microscopic structure of the $K^\pi=2^-$ state in the SLy4 model is essentially the same as in the SkM* model: 
the $\nu[622]5/2 \otimes \nu[734]9/2$ and $\pi[521]3/2 \otimes \pi[633]7/2$ excitations play a central role. 
The unperturbed energy of the $\pi[521]3/2 \otimes \pi[633]7/2$ excitation is 1.93 MeV and 1.41 MeV in $^{246}$Cm and $^{248}$Cf; 
a 0.5 MeV drop is comparable to the calculation in the SkM* model, as shown in Fig.~\ref{fig:ME_oct}. 
The reduction of the pairing of protons, associated with a strong shell closure, accounts for the drop in energy. 

As one sees in Fig.~\ref{fig:shell}, the SkM* and SLy4 functionals give different shell structures of neutrons around $N=150$: 
the single-particle orbitals with $\Omega^\pi=7/2^+$ and $9/2^-$ are inverted. 
Therefore, the $\nu[622]5/2 \otimes \nu[734]9/2$ excitation is a hole--hole type excitation in the SkM* model, 
and this becomes a particle--hole type in $N=148$.
One can thus expect that the collectivity of the $K^\pi=2^-$ state is strongest in $^{246}$Cf. 
Indeed, the SkM* model gives a strongly collective $K^\pi=2^-$ state at 0.65 MeV, with the isoscalar strength of $7.55 \times 10^5$ fm$^6$, 
which is predominantly generated by the $\nu[622]5/2 \otimes \nu[734]9/2$ excitation with a weight of 0.31 and 
the $\pi[521]3/2 \otimes \pi[633]7/2$ excitation with a weight of 0.56. 
The sum of backward-going amplitudes is 0.35, and the stiffness parameter is 218 MeV. 
If the $K^\pi=2^-$ state is observed in a future experiment at lower energy than in $^{248}$Cf, 
our understanding of the shell structure around $N=150$ will be greatly deepened. 
The ground state in $^{249}$Cf is $\nu[734]9/2$ and the excited bands of the $\nu[622]5/2$ and $\nu[624]7/2$ appear 
at 145 keV and 380 keV, respectively~\cite{abu11}---it suggests the ordering of the single-particle levels as given in the SkM* model though 
the deformed gap of $N=150$ must be larger than that of 148.
 
\subsubsection{$K^\pi=2^+$ state}\label{gam_mode}

I show in Fig.~\ref{fig:QRPA2}(b) the intrinsic excitation energies of the $K^\pi=2^+$ state in the $N=150$ isotones. 
The excitation energy calculated is higher than that of a typical $\gamma$ vibration: 
experimentally, the $2^+_2$ states were observed at low excitation energies ($\sim 800$ keV) for the well-deformed dysprosium ($Z=66$) and erbium ($Z=68$) 
isotopes around $N=98$~\cite{NNDC} and in the neutron-rich region~\cite{wat16,zha19}. 
It is noted that the present nuclear EDF method well describes the $\gamma$ vibration in the neutron-rich Dy isotopes~\cite{yos16}.

\begin{figure}[t]
\begin{center}
\includegraphics[scale=0.28]{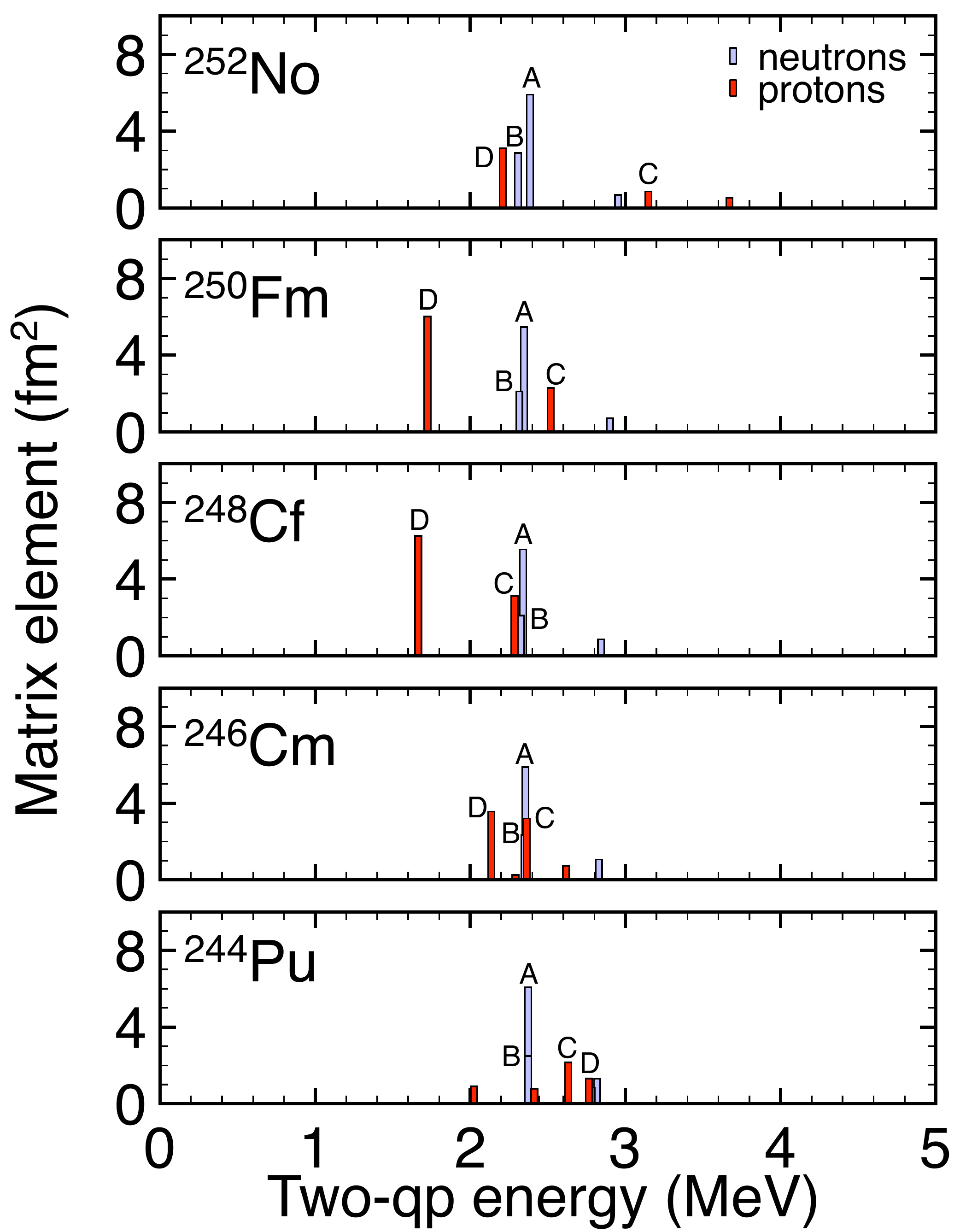}
\caption{\label{fig:ME_gam}
Similar to Fig.~\ref{fig:ME_oct} but for the $K^\pi=2^+$ state. See Tab.~\ref{tab:ME_gam2} for the configurations A, B, C, and D.
}
\end{center}
\end{figure}

One can see a similar isotonic dependence of the excitation energy to the $K^\pi=2^-$ state: a drop at around $Z=98$. 
As for the $K^\pi=2^-$ state, this is governed by the underlying shell structure. 
Let me then investigate the isotonic evolution of the microscopic structure. 
Figure~\ref{fig:ME_gam} shows the matrix elements $M_{\alpha \beta}$ defined in Eq.~(\ref{eq:ME}) of the 2qp excitations near the Fermi levels for the $K^\pi=2^+$ state. 
One sees that 2qp excitations near the Fermi levels have a coherent contribution to generate the $K^\pi=2^+$ state with the same sign as for the $K^\pi=2^-$ state. 
Table~\ref{tab:ME_gam2} summarizes the amplitudes $|f|^2-|g|^2$ for the configurations possessing a dominant contribution to the $K^\pi=2^+$ state. 
Both the neutron 2qp excitations A: $[620]1/2 \otimes [622]5/2$ and B: $[622]3/2 \otimes [624]7/2$ 
satisfy the selection rule of the $\gamma$ vibration~\cite{BM2}: 
\be
\Delta N= 0 \hspace{3pt}\mathrm{or}\hspace{3pt} 2, \Delta n_3 =0, \Delta \Lambda = \Delta \Omega= \pm 2.
\label{selection}
\ee
However, the matrix element for B is small as this configuration is a particle--particle excitation.

\begin{table}[t]
\caption{\label{tab:ME_gam2} 
Amplitudes $|f|^2-|g|^2$ of the 2qp excitations possessing a dominant contribution to the $K^\pi=2^+$ state.}
\begin{ruledtabular}
\begin{tabular}{cccccc}
configuration & $^{244}$Pu & $^{246}$Cm & $^{248}$Cf & $^{250}$Fm & $^{252}$No  \\
\hline
A: $\nu[620]1/2 \otimes \nu[622]5/2$ & 0.31 & 0.27 & 0.18 & 0.20 & 0.30 \\
B: $\nu[622]3/2 \otimes \nu[624]7/2$ & 0.14 & 0.11 & 0.07 & 0.08 & 0.20 \\
C: $\pi [521]1/2\otimes \pi[512]5/2$ & 0.12 & 0.18 & 0.14 & 0.09 & 0.03 \\
D: $\pi[521]1/2 \otimes \pi[521]3/2$ & 0.07 & 0.27 & 0.49 & 0.52 & 0.32
\end{tabular}
\end{ruledtabular}
\end{table}

The proton 2qp excitation D: $[521]1/2 \otimes [521]3/2$ governs the isotonic dependence of the $K^\pi=2^+$ state, 
that satisfies the selection rule as well. 
The $\pi[521]3/2$ orbital is located above the Fermi level of protons in $^{244}$Pu and $^{246}$Cm, 
and the $\pi[521]1/2$ orbital is located below the Fermi level in $^{252}$No, as seen in Fig.~\ref{fig:shell}. 
Thus, the transition matrix element of the 2qp excitation D is small in these isotones. 
The proton 2qp excitation C: $[521]1/2\otimes [512]5/2$ is a particle--hole excitation in $Z=94\textendash100$, 
and this has an appreciable contribution in the isotones under the present study. 
However, the particle--hole energy is relatively high, and the matrix element is not significant because the excitation is in disagreement with the selection rule. 
In this respect, $^{248}$Cf and $^{250}$Fm are the most favorable systems where the $\gamma$ vibration occurs in the low energy. 
The excitation energy of the $K^\pi=2^+$ state is higher than the $K^\pi=2^-$ state. 
This is because the quasi-proton [521]1/2 orbital is higher in energy than the [633]7/2 orbital. 
However, the SkM* model predicts that the $K^\pi=2^+$ state appears lower than the $K^\pi=2^-$ state in $^{250}$Fm; 
this may be in contradiction with the measurements of the low-lying states in $^{251}$Md~\cite{bro13}, 
where the ground state is suggested to be $\pi[514]7/2$ and 
the excited band of the $\pi[521]1/2$ configuration appears at 55 keV as the lowest state. 
 
Finally, I investigate the roles of the pairing correlations. 
Plotted in Fig.~\ref{fig:QRPA2}(b) are the results obtained by decreasing the pairing strength for neutrons (open triangles-down) and 
by increasing the pairing strength for protons (open triangles-up). 
One sees that the decrease of the pairing for neutrons lowers the excitation energies, and the effect is larger in the Pu, Cm, and No isotones. 
The increase of the pairing for protons raises the excitation energies in the Cf and Fm isotones. 
As the increase (decrease) of the pairing strength enhances (reduces) the pairing gap, the QRPA frequencies rise (decline). 
The pairing effect depends on the structure of the states. 
To see one aspect of the structure, let me introduce the quantity: $|M_\nu/M_\pi|=|\langle i|\hat{F}^\nu|0\rangle/\langle i|\hat{F}^\pi|0\rangle|$. 
For the $K^\pi=2^+$ states calculated in the SkM* model, 
this is 0.93, 0.89, 0.85, 0.86, and 0.98 in $^{244}$Pu, $^{246}$Cm, $^{248}$Cf, $^{250}$Fm, and $^{252}$No, respectively. 
Here, the $|M_\nu/M_\pi|$ value is divided by $N/Z$. 
The excitation of protons is stronger in the Cf and Fm, which is in accordance with the microscopic structure discussed above. 
The pairing of neutrons (protons) is thus relatively sensitive in the Pu, Cm, No (Cf, Fm) isotones. 
The dynamical pairing enhances the collectivity: the energy shift due to the residual pair interaction is 0.06 MeV--0.12 MeV. 
This is equal to, but more substantial than, the case in the $K^\pi=2^-$ state, indicating that the coherence among particle--hole and particle--particle (hole--hole) excitations takes place 
more strongly than in the $K^\pi=2^-$ state.

\section{Summary}\label{summary}

I have investigated the microscopic mechanism for the appearance of a low-energy $K^\pi=2^-$ state in the $N=150$ isotones in the actinides, 
in particular, a sharp drop in energy in $^{248}$Cf. 
Furthermore, I have studied the possible occurrence of the low-energy $K^\pi=2^+$ state to elucidate the mechanism 
that prefers the simultaneous breaking of the reflection and axial symmetry to the breaking of the axial symmetry alone in this mass region. 
To this end, I employed the nuclear energy-density functional (EDF) method: the Skyrme--Kohn--Sham--Bogoliubov and the quasiparticle 
random-phase approximation were used to describe the ground state and the transition to excited states. 

The Skyrme-type SkM* and SLy4 functionals reproduce the fall in the energy of the $K^\pi=2^-$ state at $Z=98$, 
where the proton two-quasiparticle (2qp) excitation $[633]7/2 \otimes [521]3/2$ plays a decisive role for the peculiar isotonic dependence. 
I have found interweaving roles by the pairing correlations of protons and the deformed shell closure at $Z=98$ formed by the [633]7/2 and [521]3/2 orbitals: 
the SLy4 model produces a strong shell closure, and accordingly, the pairing of protons is greatly suppressed, resulting in a drop in energy. 
The SkM* model predicts that the $K^\pi=2^-$ state appears lower in energy in $^{246}$Cf than in $^{248}$Cf as the Fermi level of neutrons is located 
in between the $[622]5/2$ and $[734]9/2$ orbitals. 

The $K^\pi=2^+$ state is predicted to appear higher in energy than the $K^\pi=2^-$ state 
as the quasi-proton orbital $[521]1/2$ is located above the $[633]7/2$ orbital, except for $^{250}$Fm in the SkM* model with being unlikely based on the spectroscopic study of $^{251}$Md. 
Compared with the available experimental data, the EDFs presently used are not perfect at quantitatively describing pair correlations and shell structures. 
The present study shows a need for improvements in the EDFs based on further comparative studies for describing pair correlations and shell structures in heavy nuclei, 
that are indispensable in predicting the shell effect and stability of superheavy nuclei. 

\begin{acknowledgments} 
This work was supported by the JSPS KAKENHI (Grants No. JP19K03824 and No. JP19K03872). 
The numerical calculations were performed on Yukawa-21 
at the Yukawa Institute for Theoretical Physics, Kyoto University.

\end{acknowledgments}

\bibliography{ref}

\end{document}